# MAPS OF DIPERSIONS FOR MACHINING PROCESSES

Valéry Wolff [a,*], Arnaud Lefebvre [b], Jean Renaud [c]

[a] *Université Claude Bernard – Lyon 1,
IUT B, Département Génie Mécanique et Productique,
17 rue de France, 69 627 Villeurbanne Cedex, France*

[b] *Laboratoire PRISMa, Université Claude Bernard – Lyon 1,
43 bd du 11 novembre 1918, 69 622 Villeurbanne Cedex, France*

[c] *ERPI, Equipe de Recherche des Processus Innovatifs,
8 rue Bastien Lepage, 54 360 Nancy Cedex, France*

---

**Abstract**

During the products design, the design office defines dimensional and geometrical parameters according to the use criteria and the product functionality. The manufacturing department must integrate the manufacturing and the workpiece position dispersions during the choice of tools and machines operating modes and parameters values to respect the functional constraints. In this paper, we suggest to model the turning dispersions taking into account not only geometrical specifications of position or orientation but also the experience of method actors. A representation using the principle of know-how maps in two or three dimensions is privileged. The most interesting aspect is that these maps include tacit and explicit knowledge. An experimental study realized on a machine tool (HES 300) allows to elaborate knowledge maps especially for the turning process.

*Keywords*: knowledge capitalization, know-how maps, machining dispersions, tolerancing and dimensioning.

---

## 1 Introduction

In a stronger competition context, companies widely become aware that their knowledge and their know-how constitute an important competitive advantage. In product design and manufacturing process, many authors showed the interest to save and to use the operators and the industrial experience of experts. This shows the interest to have expert knowledge ready to use [17][29][2].

Usually experience is described as being made up of two components [28]:

* Corresponding author. Tel: 04 72 65 54 80
  E-mail address: valery.wolff@iutb.univ-lyon1.fr

- On the one hand, explicit knowledge which is the type of knowledge that an individual has acquired mainly in school and university. Explicit knowledge implies factual statements such as material properties, technical information and tools characteristics. Thus explicit knowledge can be expressed in words and numbers and is therefore easily communicated and shared [16]. This knowledge is objective or unbiased.

- On the other hand, tacit knowledge which is highly personal and hard to communicate or to share with others. Tacit knowledge is deeply rooted in an individual experience and it consists in belief, and perceptions stored deep in the worldview of an individual that we take them for granted [18]. Tacit knowledge equals practical know-how. This knowledge is mostly subjective.

However the extraction of this kind of knowledge called "expert knowledge" [21][22] is not easy. We would define "expert knowledge" as knowledge that integrates not only theoretical knowledge based on known scientific or technical principles but also the expert's choice making mechanisms or behaviour as well as the decision making environment which is a fundamental factor when capitalizing on expertise. Experience is practical and not theoretical. Some methods and models are necessary to extract and formalize knowledge.

These two components of knowledge require effective and additional methods. A generic approach to knowledge capitalization is made up of three integrated phases: locating and extracting knowledge, modelling and using models.

- The first phase consists in identifying and extracting tacit knowledge from the product or process design according to the decision-maker's point of view. This involves to take measurements within the framework of an experimental strategy. This extraction of knowledge is followed by a proposition to structure the knowledge in order to understand and to model it. Measurements can be evaluated by an appropriate tool or by an expert. In both cases they are called experimental data [25][10].

- The second phase consists in modelling knowledge and confirming it. This phase involves to choose a model (a knowledge model, a behavioural one, or a hybrid) in accordance with the phenomenon being studied. The structure of the chosen model must be adapted to its function or use but the choice remains a human responsibility. Once the model has been chosen, its parameters can be adjusted by an expert or by identification using experimental data. The validation of the model will guarantee its pertinence and accuracy before it is used [23][4].

- The third phase consists in using "ready to use" knowledge. It is important to structure knowledge to reuse it. The aim is to obtain operational and long-lasting models [24][34][35].

## 2 Capitalization and cartography of the expert knowledge

### 2.1 Methods of capitalization

Several methods of capitalization exist and can be applied to build a company memory. These methods are based on the return of experiences. The most used methods are:

- The MKSM Method (Methodology for Knowledge System Management). Knowledge is modelled according to three dimensions: information, signification and context of study. Every dimension is composed of data processing, activity of the domain and tasks. This method allows to describe knowledge and to manage it [13].

- The REX Method allows to extract elements of experiences from activities and to restore them in an objective of knowledge reuse. Originally, this type of approach was applied specifically to high risk environments such as nuclear thermal power station [32][33][38].

- The MEREX Method consists in the consideration of positive and negative experiences from innovations, return of experiences, during the design of new products [6].

We propose another approach of knowledge capitalization by know-how maps introduced by ERPI and PRISMa laboratories. The interest of these know-how maps is to take into account the tacit and explicit knowledge of an expert. The aim is to re-use knowledge during the design process for any new product. These maps become of real help in the decision making process.

### 2.2 Know-how maps principle

The benefits of these maps lie in the graphic representation of the "expert knowledge" and its possibility to propose different areas of technical feasibility described according to known variations for several parameters. Theses maps allow also taking into account the processes of experts reasoning represented as areas of interest according to studied industrial priorities. Finally, they propose a traceability of product / process knowledge and a transmission of this knowledge among the various experts during time [5].

Our methodology brings another methodological answer compared with the cognitive map [15][19][36]. The objective of these cognitive maps is to represent the structure of the causal assertions of a person. The concept of modelling the cognitive process comes from psychology [11][12]. Cognitive maps allows to model in graphic representation knowledge of an individual or a group concerning a particular object. Langfield- Smith underlines that a cognitive map is not a durable structure [20][40]. It corresponds to a passing collective cognition. Cognitive maps are usually derived through interviews and so they are intended to represent the subjective world of the interviewee. Cognitive mapping is a formal modelling technique with rules for its development. Knowledge is not a data or a fact. It represents rather a network of information related to an object as shown in table 1; we suggest a comparison between two concepts of cognitive maps.

In the next section we are presenting the principle of know-how maps concerning a study of dispersions related to a turning manufacturing process.

### 2.3 Know-how maps applied to expert knowledge

The principle of know-how map consist in formalizing knowledge "ready to use" into a graphical model in a concurrent engineering context. A know-how map describes a set of expert knowledge (from the design to manufacturing processes) and is represented in a graphical form. The construction of these know-how maps includes three main stages: extraction, modelling and using knowledge. Our objective aims at showing at the "t" moment the expertise of one or several individuals for a given environment and a given activity or operational task. This involves managing the individual knowledge of the decision maker in order to formalize it and use it in a collective way. Expert knowledge is extracted and structured modelled, and applied to improve the design process. Some authors have focused on knowledge capitalization at the intersection between two disciplines: knowledge engineering and human management. The knowledge map remains incomplete. It corresponds to a representation of a field of activity according to the point of view of the experts. The know-how maps include two sorts of knowledge [1]:

- O*bjective knowledge or knowledge engineering*. Knowledge is general, not connected to a precise problem. It is relative to the function of a system and to the causal relations between the system variables. Explicit knowledge is represented by mathematical models.

- *Subjective knowledge or tacit knowledge*. This knowledge is formalized by heuristic forms which reflect the experience of the experts. They are specific in the treated problem and the expert who designed the system. Expert's rules or reasoning schemes are often used in these maps.

Identify an expert remains a difficult task. Shanteau [31] proposed nine experts' levels (experience, certification, social cheer, consistency reliability, consensus reliability, discrimination, behavioural characteristics….). He also proposed a tool to estimate an expert according to two indications of discrimination (large variety) and logic (repetition).

The methodology of know-how maps consists in representing under a 2D graph a response function depending on continuous parameters. The three main steps of modelling are:

*Identification of knowledge*: this step consists in identifying various product / process parameters of the studied system (manufacturing features …).

*Modelling of explicit knowledge and tacit knowledge*: the maps construction includes three parts:
- the first part aims at determine the equation of regression obtained for example by a design of experiments.
- the second part aims at the research of feasibility areas according to the expert's knowledge.
- the third step consists in giving production rules depending on the various working areas.

*Use of the know-how maps*: the know-how maps are used during the product/process design.

Figure 1 and figure 2 show an example of know-how maps construction which represents the evolution of a process parameter $P_k$ (number of manufacturing operations) according to two products parameters $P_i$ and $P_j$. Feasibility areas 1 and 2 (figure 1) are defined by a mathematical model (explicit knowledge). According to the expert knowledge, a third area is defined (figure 2). Three areas are then defined and correspond to manufacturing operations for a given manufacturing feature.

### 2.4 Benefit of know how maps

The know-how mapping presented in this paper is a methodological tool to collect information and knowledge of experts minds. This tool allows to propose "ready to use" knowledge for the industrial decision-maker. The representation of knowledge as know-how maps presents several interests:
- it allows to include practices and experiences of the experts and it allows to make them understandable [7][8],
- it allows to take into account processes of reasoning and thought of experts under feasibility areas according to studied industrial priorities,
- it can be used for educational finality. It becomes a support in the discussion and exchange among experts. The know how maps become a tool of visual communication, a real practical guide for the decision-maker,
- it allows to assure the transmission of knowledge among the various experts during time,
- it allows to integrate objective and subjective knowledge on the same graphic support.
Finally, it allows an update of the expertise according to the evolution of products and used practices. However, the know-how map remains partial and need expert's rules to improve the model. The map must be constantly put back about the evolution the expert. It corresponds to a representation at a level of study according to experts point of view.

### 2.5 Industrial applications

Knowledge representation in the form of know-how maps using recommendations responds to a strong demand from people involved in design and manufacturing. Maps construction depends on the specialist point of view who is in charge of its elaboration. But these maps have to be generic enough to represent product and associated processes in the concurrent engineering context.

Several industrial studies about know-how maps elaboration have been achieved around different manufacturing context since about ten years: food supplying industry, halogen lamps manufacturing within the Philips Company [41] and camshaft bearing lines machining on five axis centers [21] within Renault company.

In the next section, we present an application of the know-how maps to study manufacturing dispersions related to a turning process.

## 3 Know-how maps applied to machining dispersions in turning

The objective of this third part consists in developing the design process of know-how maps starting from a study of machining dispersions. The evaluation of dispersions allows the development of know-how maps related to a type of machine-tool and a standard part fixture.
Calculations of manufacturing dimensions starting from the geometrical definition of the part can be generally obtained from two methods [3][30]: the installation of chains of dimensions or the method of dispersions. It is this second method which we adopted in this study.

### 3.1 Dispersions modelling

We call machining dispersions [9] the geometrical and dimensional variations obtained on a series of real parts for a manufacturing process and a given machine-tool. The supposed sources of dispersions have several origins in particular related to controls with the inflexion of the tools, the cutting efforts and the geometrical defects of the machine tool. The rule which characterise each origin of dispersions can be of various mathematical models (Normal or Poisson distribution…); nevertheless we make the assumption that the resultant response follows a Normal distribution [27].

The industrial need is to control each machine according to its machining dispersions in time. The model we propose can fit to a range of machine-tool (e.g.: turning, milling …) but each machine has its own characteristics and then its own dispersions values.
Machine characteristics (slides clearance, drive systems …) for a given machining operation can vary in time. Thanks to the experiments limited numbers proposed by our model, it is possible to update periodically the dispersions values and then to optimize the manufacturing process.

#### 3.1.1 Extended proposed model

Considering a shouldered part, the modelling of the behaviour of a lathe is classically approached according to a thorough study of five parameters of dispersions ($\Delta_{machine}$) as shown in the figure 3. These dispersions are classified into two categories: the first relates to dispersions of setting in position (remachining) such as $\Delta O$, $\Delta \alpha$, $\Delta Z_r$. The second category includes machining dispersions like $\Delta R_u$ (dispersion of machining according to X axis) and $\Delta Z_u$ (dispersion of machining according to the Z axis: spindle axis).
For the taking into account of the axial dimensions and the dispersions obtained during machining, we use the traditional method of $\Delta L$ on X and Z axes. The parameters are then indicated by $\Delta Z_u$, $\Delta R_u$, $\Delta Z_r$. The taking into account of the geometrical specifications [39] (coaxiality, perpendicularity…) involves the use of new parameters $\Delta \alpha$ and $\Delta O$.

- $\Delta \alpha$ represents the angular remachining error of the part in the soft jaws in turning,

- $\Delta O$ is the defect of concentricity (between the axis of the reference surface and the spindle axis) located at the bottom of the soft jaws,

- $\Delta Z_r$ corresponds to the axial remachining error of the part in the part holder along the Z axis.

The objective of the proposed model is to determine the relationships between the machine-tool parameters and the product/process parameters.

These relations are of the type:

$$\Delta_{machine} = f_i(p_i)$$

where $\Delta_{machine}$ is one of five dispersions and $p_i$ a set of parameters (discrete or continuous).

### 3.1.2 Dimensioning according to ISO standards

Standards ISO of dimensioning and tolerancing, gathered under the term of GPS (Geometrical specification of the Products) provide a complete language to mechanical engineers. They are adopted today by the manufacturing industry. The geometrical model of dispersions in simulation of machining which we propose takes into account these three dimensional specifications.

For example, in the case of the coaxiality, we defined the methods of calculation necessary to connect the model parameters to ISO specifications.

The coaxiality relates to the relative position of the real axis of specified surface and the reference/datum axis. It never relates to surfaces but always to axes. The definition resulting from the standard and its interpretation is defined in the figure 4.

This definition must be interpreted and calculated to make corresponding the obtained measurement on coordinate measuring machine (CMM) with the part design specifications.

## 3.2 Representation of the dispersions model: maps from experimental design

We now propose to extend the field of application of the model. It must take into account the variations of certain parameters of the manufacturing process likely to involve variations of dispersions. The method of the experimental design is exploited here to quantify the influence of these modifications on dispersions.

To quantify the relations $f_i$ between dispersions and product/process parameters, we use the design of experiments methodology [14][37][26]. Indeed, the described knowledge by the $f_i$ functions comes under the tacit and behavioural field of major knowledge. The design of experiments as well as the general models of linear regressions is well adapted to the determination of the $f_i$ functions.

We call maps of dispersions related to expert knowledge a chart of a dispersion function $y = f(x_i)$ into two or three dimensions, according to product or process parameters resulting from a formal modelling.

In order to use these maps of dispersions in a predictive mode, we must obtain a representation in two dimensions which involves that the model contains a maximum of two continuous parameters among the $x_i$.

For each combination of discrete parameters, we vary simultaneously two continuous parameters $x_1$ and $x_2$ into a range of variation specified by the experts. Then, we calculate the theoretical response. We represent each function for a combination of discrete parameters. Moreover, for correct use, criteria of use must be added on the know-how maps.

## 3.3 Experimental protocol

### 3.3.1 Design of experiments parameters

We defined five product/process parameters to evaluate various dispersions. The table 2 gives the list of the parameters. Some particular interactions are taken into account (see table 3).

### 3.3.2 Determination of the design of experiments

We choose the Taguchi's method to limit the number of tests to take into account times of machining and control. Taking into account the criterion of orthogonality and number of freedom degrees, the $L_{16}$ ($2^{15}$) table was selected. For each response of studied dispersion, the order of the tests as well as the combinations of the parameters are given in table 4.

### 3.4 Experimental results

#### 3.4.1 Evaluation of responses

Responses of the experiments are obtained by measuring. Measures allow calculating the dispersions parameters $\Delta O$, $\Delta\alpha$, $\Delta R_u$, $\Delta Z_r$ and $\Delta Z_u$ (in millimetre). A reduced sample of five workpieces is sufficient. The Taguchi's method used is a standard $L_{16}$ ($2^{15}$) experiment, 5 times repeated.

We observe two cases. The response is calculable starting from the standard deviation obtained by the measurement of one or several dimensions:

- Case of the response obtained by the measurement of only one dimension. That relates to dispersions $\Delta O$, $\Delta\alpha$, $\Delta R_u$. For example, the relation relating to $\Delta R_u$ is written:

$$\Delta Ru = \left[\frac{6 \times (\sigma_{diametral})_{sample}}{C_4}\right]/2$$

where $C_4$ is the weighting taken in the statistical table of the reduced samples.

- Case of the response obtained by the measurement of several dimensions $d_i$ (standard deviation noted $\sigma_i$). The variance of required dispersion is related to the sum of the variances of concerned dimensions. That relates to $\Delta Z_r$ dispersions (or $\Delta Z_u$). The relations used are as follows:

$$\sigma_r = \frac{\sqrt{\sum \pm \sigma_i^2}}{\sqrt{2}} \quad \text{and thus}$$

$$\Delta Zr = \frac{(\sigma_r)_{sample}}{C_4} \times 6$$

The necessary values to calculate the dispersion parameters (part diameters, points of intersection, etc…) are measured directly on the 80 parts using a coordinate measuring machine. Some of these values are only intermediates parameters.

Each batch of 5 parts allows to calculate the standard deviation of each answer $\Delta Z_r$, $\Delta O$, $\Delta\alpha$, $\Delta R_u$ and $\Delta Z_u$ using CMM measurements. The design of experiments provides in this way 16 values for each studied response.

#### 3.4.2 Analyse and summary of the obtained results

We carried out the analysis of the measurements obtained on the 80 parts of the $L_{16}$ ($2^{15}$) plan definite previously to determine the parameters of manufacture process planning influencing dispersions of machining.

The variance analysis indicates that a parameter is statistically significant on the response as soon as $p$ parameter is higher than 0.05 (Level of confidence higher than 95%).

The R-squared ($R^2$) makes it possible to evaluate the percentage of data explained by the model. The higher the $R^2$ is, the more the model is usable in a predictive mode. A coefficient $R^2$ between ~70 and ~ 90% corresponds to an acceptable model.

Table 5 presents the synthesis of the Pareto chart associated with the dispersions obtained by experimentation. For each studied answer ($\Delta Z_r$, $\Delta O$, $\Delta\alpha$, $\Delta R_u$ and $\Delta Z_u$), the diagram highlights the influence (significant or not) of the various parameters of the design of experiments on the studied response.

### 3.5 Know-how mapping for the $\Delta R_u$ response

#### 3.5.1 Map's construction

We choose to develop an example of map related to the response of $\Delta R_u$ dispersion which corresponds to the machining dispersion according to X axis.

In our study, the $f_i$ function to be represented as a chart of dispersion is thus:

$$\Delta R_u = f (\text{insert type, nose radius, material, cutting speed, feed rate})$$

The linear model of regression present in the form of:

$$\Delta Ru = \alpha + \beta N + \chi R_\varepsilon + \delta Vc + \varepsilon M + \phi f + \gamma NV - \eta Nf - \varphi VcM + \kappa Mf$$

where the coefficients are:

| $\alpha$ | 0.02789 | $\beta$ | 0.01822 | $\chi$ | 0.00637 | $\Delta$ | 0.00008 | $\varepsilon$ | 0.01187 |
|---|---|---|---|---|---|---|---|---|---|
| $\phi$ | 0.01941 | $\gamma$ | 0.00004 | $\eta$ | 0.02873 | $\varphi$ | 0.00010 | $\kappa$ | 0.06160 |

Only two parameters are continuous parameters: $V_c$ (cutting speed) and $f$ (feed rate). The representation in two dimensions is thus possible.

Each combination of discrete parameters N (insert type), $R_\varepsilon$ (nose radius), and M (m*aterial*) corresponds a $f_i$ function. The map of dispersions obtained for the combination N $R_\varepsilon$ M = (- 1-1-1) is represented by figure 5. It is a representation in two dimensions where the greyed areas correspond to the range of variation of the $\Delta R_u$ response.

#### 3.5.2 Criteria of use for the $\Delta R_u$ know-how map

For each combination of discrete parameters N $R_\varepsilon$ M, we observe that the regression equation (1) is:

$$\Delta Ru = C_1 + \alpha_1 x_1 + \beta_1 x_2 \qquad (1)$$

*$C_1$, $\alpha_1$ and $\beta_1$ are constants and $x_1$ and $x_2$ are the continuous parameters $V_c$ and f.*

The regression equation (1) is a datum plane equation. The two main areas of dispersions are a "0.02-0.04" area and a "0.04-0.06" area. The border between the two areas is the place where $\Delta R_u$ is equal to 0.04. The equation (1) becomes then: $0.04 = C_1 + \alpha_1 x_1 + \beta_1 x_2$. This is a straight line equation (figure 5). We thus consider that the points located close and on the common line are included in a third zone of dispersions "0.04" whose amplitude is defined by the expert.

When the amplitude of the interval in which the response varies is higher or equal to 0.01, we retain the value of the smallest hundredth millimetre included in this interval. For example, as shown in figure 6, the amplitude of the variation interval is: $0.06 - 0.02 = 0.04 > 0.01$. We thus identify two ranges of response: a "0.02-0.04" range and a "0.04 - 0.06" range. For the first range, the selected dispersion will be 0.02 mm and for the second range, the selected dispersion will be thus 0.04 mm.

Three areas will thus be defined by the expert: for the first, dispersion selected will be 0.02 mm, for the second, dispersion selected will be thus 0.04 mm and for the third, dispersion will be 0.06 mm (figure 6).

## 4  Know-how maps use

We present a simple application to illustrate the use of our dispersions model in the approach of know-how maps.

The study's aim is to select the best parameters to optimize the process plan.

Continuing the example of the $\Delta R_u$ which corresponds to the machining dispersions according to X axis (§ 3.5) figure 7 shows the $\Delta R_u$ values for two cutting parameters ($V_c$, $f$). These parameters are statistically significant parameters that have been highlighted by the design of experiments (§ 3.4.2).

The process planner must choose $V_c$ according to roughness and specified tolerances. Roughness allows a range of the feed rate ($f$) for each $R_\varepsilon$ value. The know-how map gives the possibility to find the higher value for $V_c$ which corresponds to an economical criteria regarding the dimensioning tolerances.

The maximal feed rate $f$ is 0.15 mm/tr for a roughness $R_a$ equal to 1.6 µm on a turned workpiece with $R\varepsilon$ equal to 0.4 mm (Sandvik documentation). As shown on figure 7 dispersions increase with the cutting speed $V_c$. Then for a turned diameter $\varnothing$ 50 H8 the maximum cutting speed allowed to respect the dimensioning tolerances is equal to 200 m/min (figure 6).

## 5  Conclusion

One of the key of the knowledge capitalisation process is the expert know-how re-use for new products development in an innovation context.

In this article we present several methods used in the industry and we propose a new approach which consists in knowledge modelling by know-how maps. These maps are 2D graphical representations of experts' knowledge and are associated with technical recommendations. The main advantages of these "ready to use maps" are the time decrease of the design and the manufacturing processes and the optimisation of manufacturing parameters.

The control of the manufacturing dispersions promotes the decrease of the gap between the functional "target" product and the real manufactured product. Moreover the control operations allow to quantifying it. The approach developed in this article aims at spreading the classic dispersions model by taking into account the geometrical specifications of orientation and position. Moreover, the formalization of information exchanged between design and manufacturing offices is one of the key factors for the decrease of the time necessary to design product.

The know-how maps allow the experts to take into account the influence of manufacturing parameters variations (cutting parameters and equipment used) on machining dispersions. In addition to the predictive aspect, the user has the possibility of visualising capitalised know-how graphically and thus can better take into account capability real machine-tool according to the process considered.

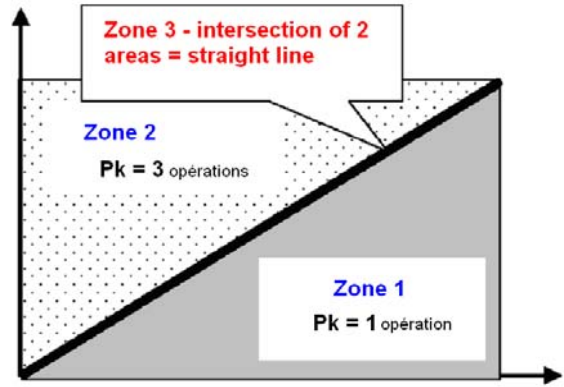

Figure 1 : Know-how map for explicit knowledge obtained with a mathematical model. Number of operations $P_k$ necessary for a manufacturing feature represented with 2 areas.

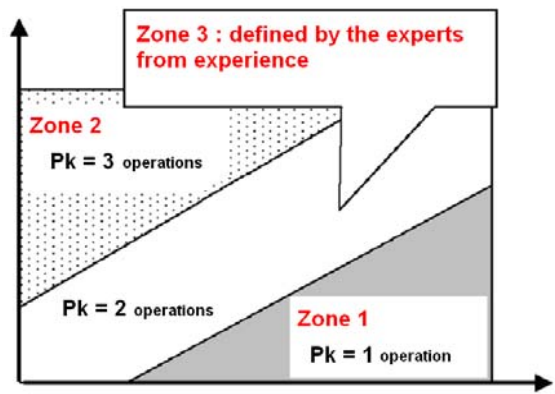

Figure 2 : Know-how map with tacit knowledge added by the experts. Expert's knowledge allows to build the third area.

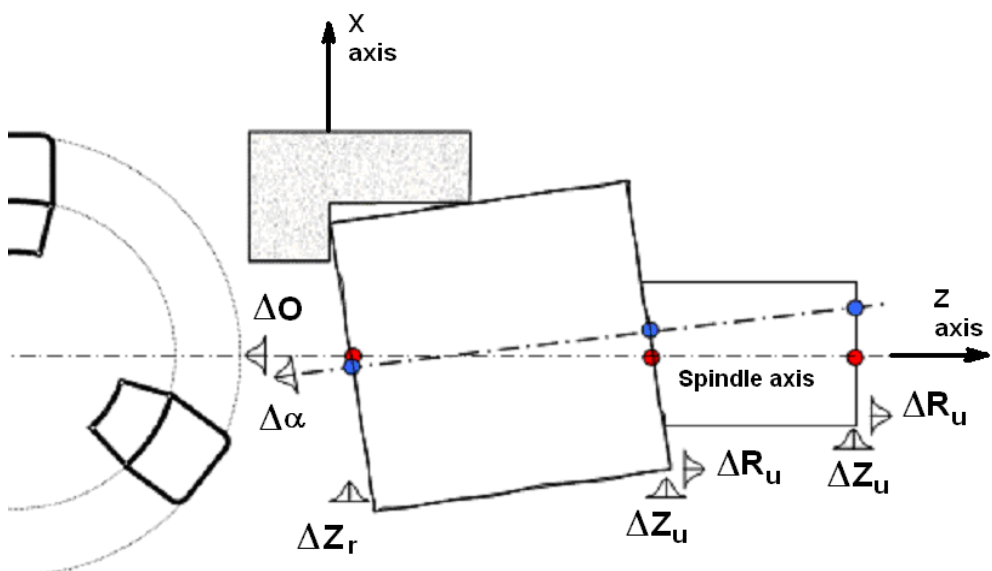

Figure 3 : Modelling of the 5 dispersions retained in turning

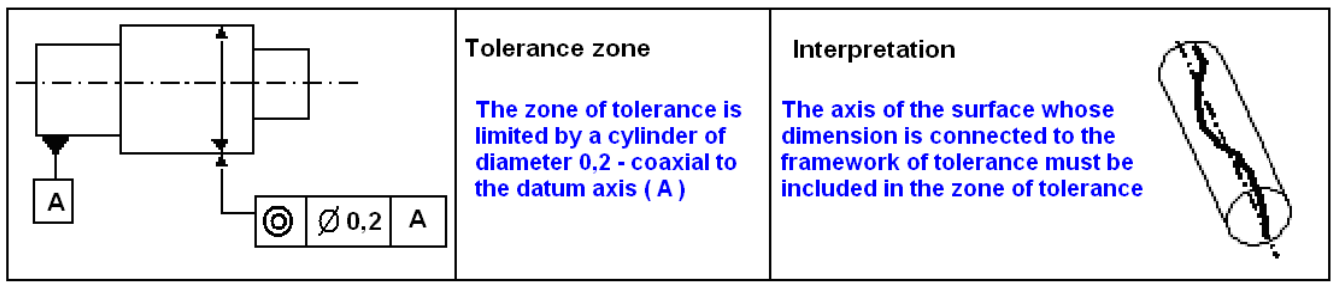

Figure 4 : GPS standards- coaxiality according to ISO 8015

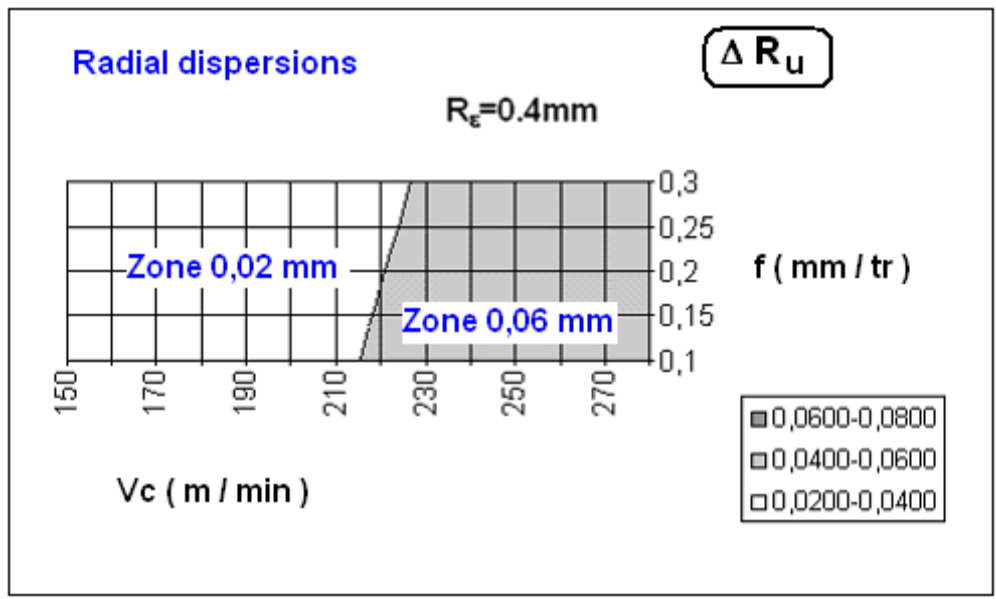

Figure 5 : map of dispersions for "P15" tool insert type, a 0.4 mm nose radius and A60 material.

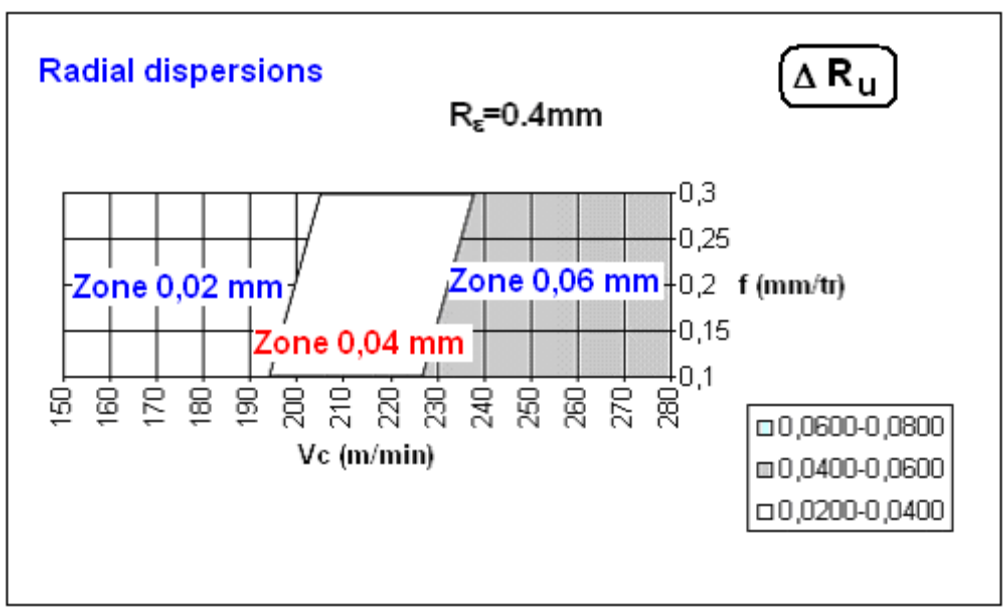

Figure 6 : definition of the various areas of know-how

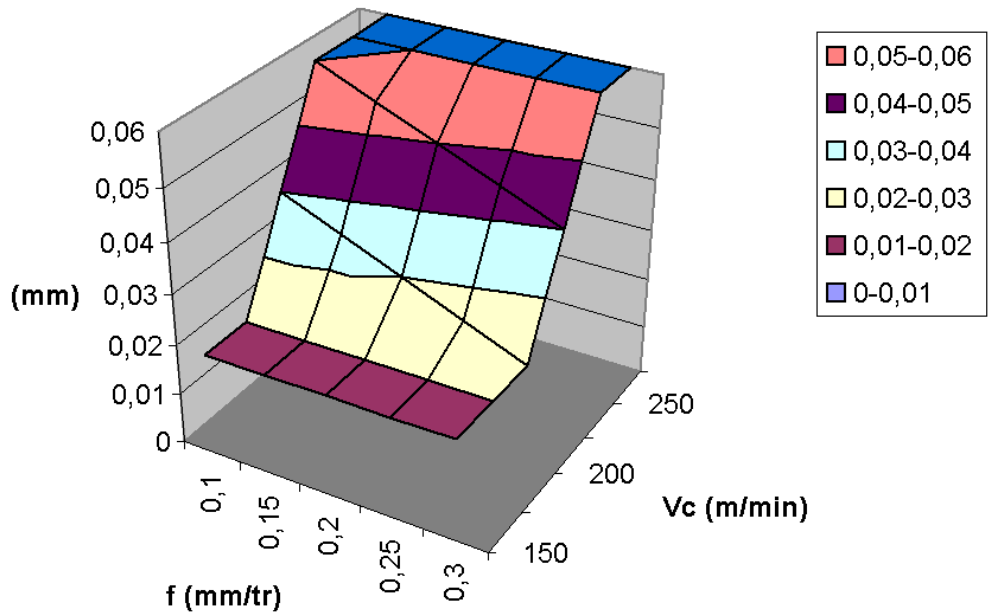

Figure 7 : dispersions evolution for $R_\varepsilon$ = 0.4 mm

|  | Cognitive Map | Know-how map |
|---|---|---|
| Convergent points | Physical support. Visual charts<br>Builds itself partly or entirely starting from the cognition of the expert<br>Integration of fuzzy or related logic | |
| Divergent points | Networks (arcs and nodes)<br>Tacit Knowledge<br>Determinist | Zone preferably<br>Taking into account of explicit and tacit knowledge<br>Choice is left to the expert |
| Advantages | Use and easy comprehension of the charts | Easy construction of tacit knowledge |
| Disadvantages | Difficult to represent<br>Complex graphs<br>Need for carrying out several front intermediate graphs to obtain the final graph | The representation remains to two or three dimensions<br>Seek mathematics models for explicit knowledge |

Table 1 : Comparison enters the cognitive charts and maps of know-how

| Parameters | | Type | Values | |
|---|---|---|---|---|
| Insert type | N | discrete | P15 | P35 |
| Nose radius (mm) | $R_\varepsilon$ | discrete | 0.40 | 0.80 |
| Cutting speed (m/min) | $V_c$ | continuous | 150 | 280 |
| Material of the machined part | M | discrete | A60 | XC38 |
| Feed rate (mm/turn) | $f$ | continuous | 0.10 | 0.30 |

Table 2 : Factors and associated values

| Interactions | |
|---|---|
| Insert type - cutting speed | $N.V_c$ |
| Material – feed rate | $M.f$ |
| Cutting speed - material | $V_c.M$ |
| Insert type – feed rate | $N.f$ |

Table 3 : Interactions retained between the parameters by the experts

| N° | Insert type | Nose radius $R_\varepsilon$ | Cutting speed $V_c$ | Material M | Feed rate $f$ |
|---|---|---|---|---|---|
| 1 | P15 | 0.4 | 150 | A60 | 0.1 |
| 2 | P15 | 0.4 | 150 | XC38 | 0.3 |
| 3 | P15 | 0.4 | 280 | A60 | 0.3 |
| … | … | … | … | … | … |
| 16 | P35 | 0.8 | 280 | XC38 | 0.1 |

Table 4 : Table of the experiments (extract)

| PARETO charts | | Variance analysis | |
|---|---|---|---|
| | | R-squared statistic $R^2$ | influential parameters ($p_i$) $p > 0.05$ |
| $\Delta O$ | 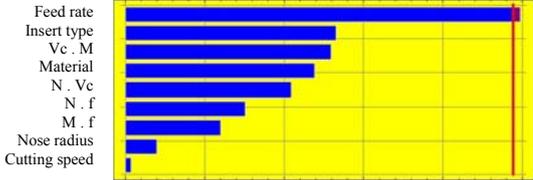 Feed rate / Insert type / Vc . M / Material / N . Vc / N . f / M . f / Nose radius / Cutting speed | 0,68 | Feed rate ($f$) |
| $\Delta \alpha$ | 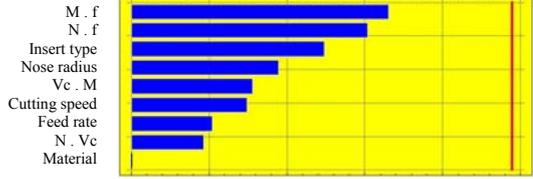 M . f / N . f / Insert type / Nose radius / Vc . M / Cutting speed / Feed rate / N . Vc / Material | 0,60 | |
| $\Delta R_u$ | 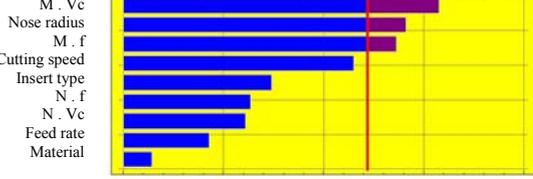 M . Vc / Nose radius / M . f / Cutting speed / Insert type / N . f / N . Vc / Feed rate / Material | 0,86 | Nose radius ($R_\varepsilon$) Material – Cutting speed ($M.V_c$) Material – Feed rate ($M.f$) |
| $\Delta Z_r$ | 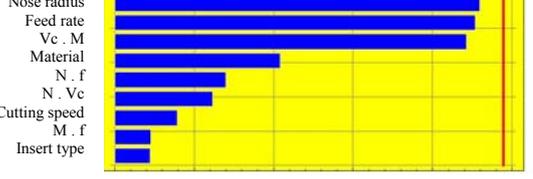 Nose radius / Feed rate / Vc . M / Material / N . f / N . Vc / Cutting speed / M . f / Insert type | 0,74 | |
| $\Delta Z_u$ | 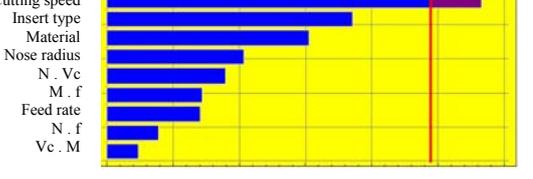 Cutting speed / Insert type / Material / Nose radius / N . Vc / M . f / Feed rate / N . f / Vc . M | 0,74 | Cutting speed ($V_c$) |

Table 5 : Synthesis of the parameters influence